\documentclass[a4paper,UKenglish]{lipics-v2018}

\usepackage{graphicx,amsmath,amssymb,url,xspace}
\usepackage{booktabs} 
\usepackage{lmodern}
\usepackage{cleveref}
\usepackage{dirtytalk}
\usepackage{array}   
\newcolumntype{L}{>{$}l<{$}} 
\newcolumntype{R}{>{$}r<{$}} 
\newcolumntype{C}{>{$}c<{$}} 

\usepackage[frozencache]{minted}
\usepackage{subcaption}
\usepackage{xcolor}

\usepackage{empheq}

\usepackage[utf8]{inputenc}
\usepackage[T1]{fontenc}
\usepackage{microtype}
\title{Proof-of-Concept Examples of Performance-Transparent Programming Models}
\author{Benjamin Andreassen Bjørnseth}{Norwegian University of Science and Technology\\Department of Computer Science\\{[Trondheim, Norway]}}{benjambj@ntnu.no}{}{}
\author{Jan Christian Meyer}{Norwegian University of Science and Technology\\Department of Computer Science\\{[Trondheim, Norway]}}{jan.christian.meyer@ntnu.no}{}{}
\author{Lasse Natvig}{Norwegian University of Science and Technology\\Department of Computer Science\\{[Trondheim, Norway]}}{lasse.natvig@ntnu.no}{}{}

\authorrunning{B.A. Bjørnseth and J.C. Meyer and L. Natvig}
\Copyright{B.A. Bjørnseth and J.C. Meyer and L. Natvig}
\subjclass{Software and its engineering$\rightarrow$ Software notations and
  tools $\rightarrow$ General programming languages}
\nolinenumbers 
\hideLIPIcs  

\begin{document}
\maketitle

\begin{abstract}

  Machine-specific optimizations command the machine to behave in a specific
  way. As current programming models largely leave machine details unexposed,
  they cannot accommodate direct encoding of such commands. In previous work we
  have proposed the design of performance-transparent programming models to
  facilitate this use-case; this report contains proof-of-concept examples of
  such programming models. We demonstrate how programming model abstractions may
  reveal the memory footprint, vector unit utilization and data reuse of an
  application, with prediction accuracy ranging from 0 to 25 \%.


\end{abstract}

\section{Introduction}
\label{sec:introduction}

When machine-specific optimizations are written by hand, the program is intended
to specify how the machine is to be used. However, current programming models
are designed to abstract away machine details; as such, they do not accommodate
direct encoding of machine-specific intentions. This complicates programming,
since the programmer is forced to ``program around the language''
\cite{Snyder1986}. The resulting programs will also only indirectly encode the
machine-specific intentions, thereby making them harder to read and appreciate.
As tools for writing machine-specific optimizations, current programming models
are therefore wanting \cite{Hoare1974}.

To address this, we have proposed the study of \textit{performance-transparent
  programming models} \cite{cfpaper}: programming models which reveal the
performance of code written in it, thus facilitating recognition of efficient
and inefficient code. We define transparency to be the capacity the programming
model has for revealing program properties, and propose measuring transparency
by the accuracy with which a program property may be derived using only
explicit, context-independent rules. Explicitness and context-independence
foster simplicity and compositionality, which are features also highlighted as
the key to the perceived descriptive power of declarative languages
\cite{van2004concepts}. To define explicitness and context-independence
precisely, we propose the restrictions that models of program properties be
expressed as S-attributed attribute grammars \cite{Paakki1995}. As a concrete
strategy to make performance transparency work, we propose designing programming
model abstractions which expose the hardware components which most severely
affects performance.

In this paper, we present proof-of-concept examples demonstrating how some
example programs which exhibit various machine-related performance effects may
be written and analysed using a hypothetical programming model syntax. For each
program, we present a performance model, and demonstrate that the model reflects
the observed behaviour of a corresponding implementation program. Our primary
purpose is to concretize the theoretical foundation; we therefore only present
models and measurements targeting one specific architecture, namely the first
generation Intel Xeon Phi processor \cite{Jeffers2013}. The processor is
suitable as an initial target due to the relative simplicity of its core design.
To focus on specific performance effects the examples are run single-threaded
with prefetching disabled, using data sets which fit and are pre-loaded into the
L2 cache.

\section{Programming Model Examples}
\label{sec:examples}




\subsection{AoS vs. SoA}
\label{sec:aos-vs.-soa}

\begin{figure}[t]
  \centering
  \includegraphics[scale=0.7]{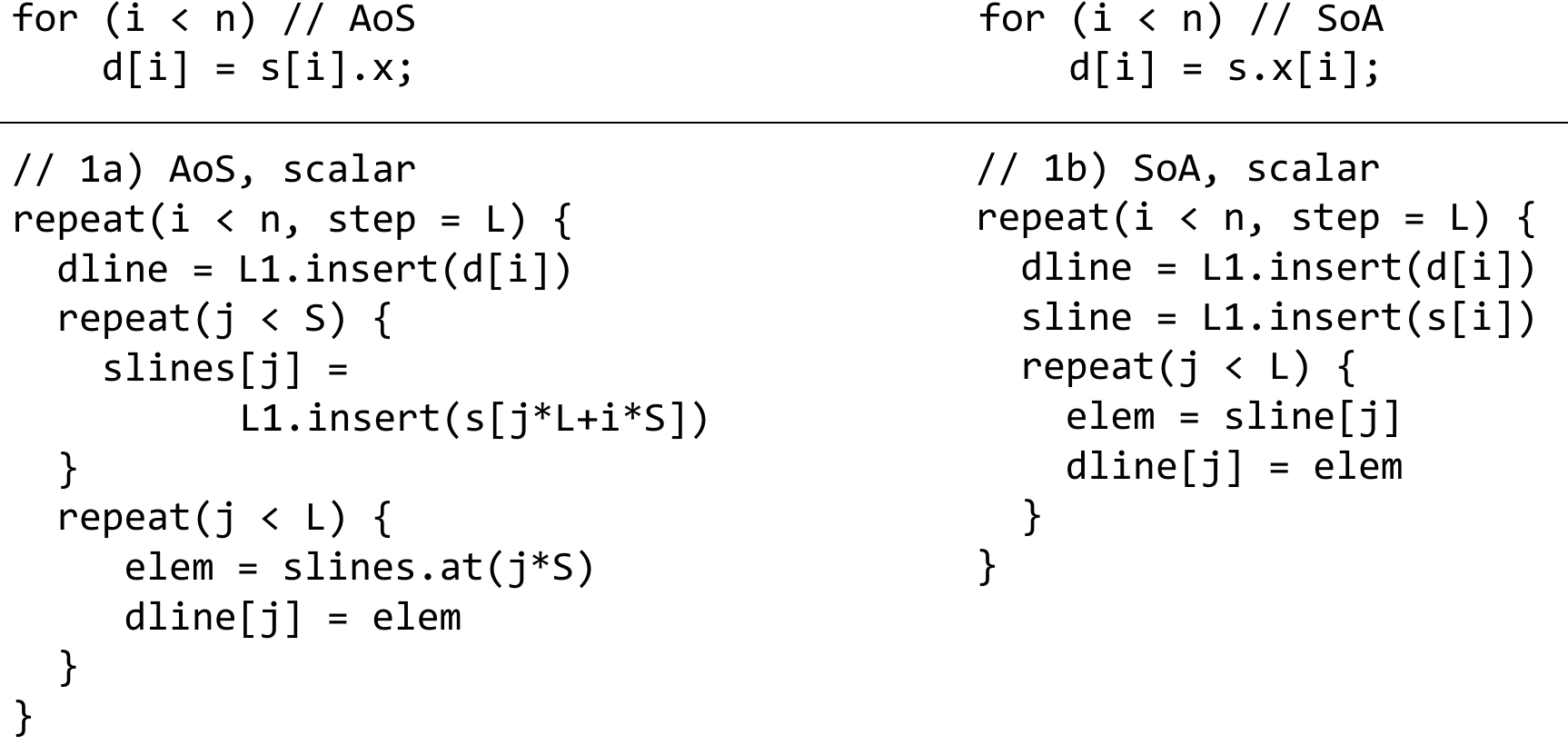}
  \caption{\label{fig:aos-vs-soa-scalar-src} Two kernels, and their
    implementation in a hypothetical transparent programming model.}
\end{figure}
%

The first example analyses two variations of a copy kernel, illustrated in the
top of \Cref{fig:aos-vs-soa-scalar-src}. The two versions both copy \texttt{n}
elements from one location in memory to another, but differ in their choice of
source vector memory layout: ``array-of-structures'' (AoS) vs.
``structure-of-arrays'' (SoA).
The best layout is
situation-specific. In this section, we will see how a performance-transparent
programming model may reveal which choice is better by highlighting memory
footprint and vector unit utilization for this particular example.

\subsubsection{Memory Footprint}
\label{sec:memory-footprint}

First, we consider a programming model which exposes the L1 data cache with a
single operation: \texttt{L1.insert(elem)}. The operation inserts the cache
block containing the specified data element into the L1 cache, and returns a
value representing the cache line with the block loaded. Only such cache line
objects can be used as memory operands to CPU instructions; thus, the type
system is leveraged to mimic the ``cold-miss'' semantic rule that data must be
inserted into the cache before use. The cache line may also be indexed to
represent specific elements in the line. The programming model also has a type
\texttt{L1Lines} which represents a sequence of contiguous L1 lines, where
individual lines can be referred to using indexing
(\texttt{lines[i]}) and individual elements can be referred to with a function call
(\texttt{at(i)}). Finally, the programming model contains a \texttt{repeat(n,
  step) body}-construct, which executes \texttt{body} $\frac{n}{step}$ times.

Using this programming model, the AoS-code can be implemented as shown in code
section 1a) in \Cref{fig:aos-vs-soa-scalar-src}. The cache line length in
elements is denoted \texttt{L}, and for simplicity we assume that \texttt{L}
divides \texttt{n}. The program loops over the index space in steps of
\texttt{L}, and write-allocates the next block of destination data to fill. It
then loads the cache blocks of source data required to copy the next \texttt{L}
elements into the destination line. However, the \texttt{x}-elements of the
source vector are separated by the other fields of the struct. If we let
\texttt{S} be \texttt{sizeof(s[i])/sizeof(s[i].x)}, then the next \texttt{L}
elements will be located on the next \texttt{S} blocks \cite{Bjornseth2017}. The
program therefore loads \texttt{S} blocks of the source vector into L1 cache.
Finally, the \texttt{L} elements are copied from the source cache lines to the
destination cache line. The SoA-code can be implemented similarly, as
demonstrated in code section 1b) of \Cref{fig:aos-vs-soa-scalar-src}. Since the
source elements are contiguous in the source vector, it suffices to insert a
single cache block from the source vector in each iteration.

\subsubsection{Vectorization}
\label{sec:vectorization}


The kernels in \Cref{fig:aos-vs-soa-scalar-src} can also be implemented using
vector instructions to copy more than one element at a time. In a
non-transparent language such as C, one cannot in general know whether vector
instructions will be used: this is compiler-specific, and depends both on the
compiler heuristics and on how sensitive it is to the way the source code is
written. Several programming models have therefore been built with the purpose
of taking explicit control over code vectorization \cite{Wang2009a, Leissa2014,
  Esterie2014a, Karpi2017, Fog2013, intel-intrinsics}. To focus on performance
transparency, this section presents a simple, example-specific, hypothetical
programming model designed to be performance-transparent.

\begin{figure}[t]
  \centering
  \includegraphics[scale=0.7]{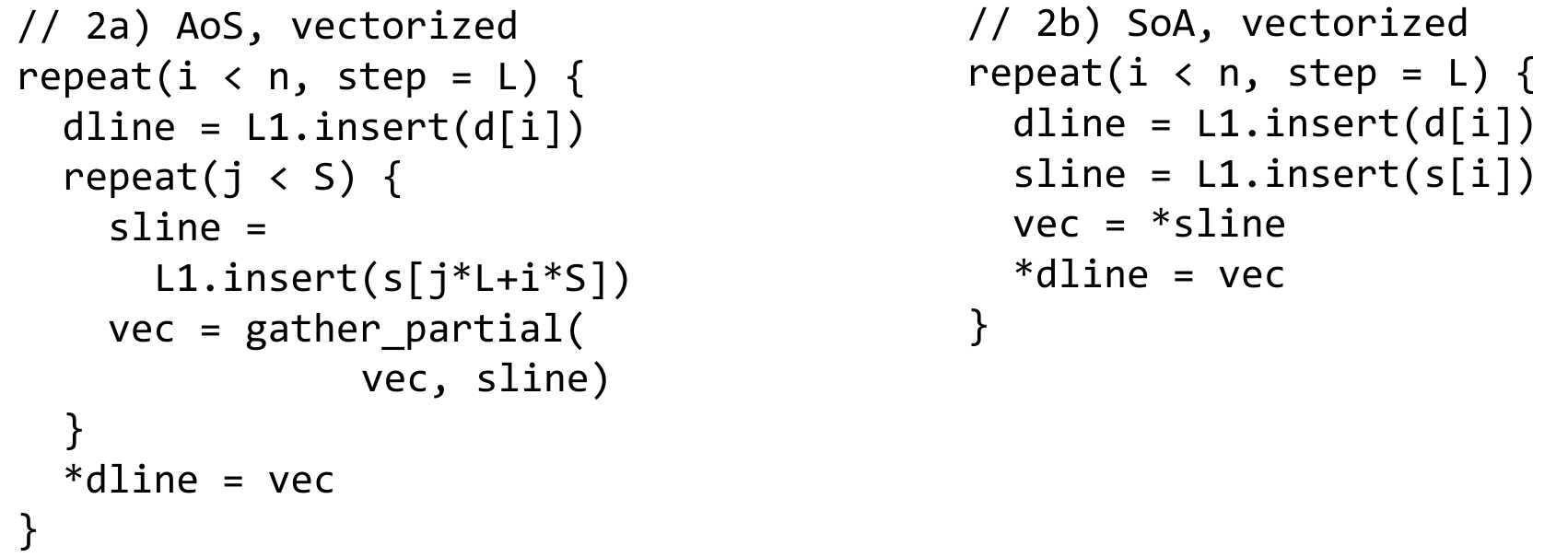}
  \caption{\label{fig:aos-vs-soa-vec-src} Vectorized implementations of the code
    in \Cref{fig:aos-vs-soa-scalar-src}.}
\end{figure}

\Cref{fig:aos-vs-soa-vec-src} demonstrates how we may represent the vector unit
behaviour. Vector loads and stores are represented by dereferencing the cache
line variables returned by \texttt{L1.insert()}, since the vector size in the
Xeon Phi coprocessor equals the cache line size. If the vector to be loaded
consists of elements scattered over multiple cache lines, the load must be
implemented using a separate \texttt{gather()} instruction. However, the
instruction as implemented on the Xeon Phi coprocessor must be executed in a
loop as it only fetches the data stored in one cache line at a time
\cite{Hofmann2014}. The programming model therefore represents the instruction
with a \texttt{gather\_partial(vec, line)} construct, which stores the relevant
elements from cache line \texttt{line} in the vector \texttt{vec}. For
simplicity, we do not represent the mask management used to determine which are
the relevant elements in the current cache line.

\subsection{Broadcast}
\label{sec:broadcast}

\begin{figure}[t]
  \centering
  \includegraphics[scale=0.70]{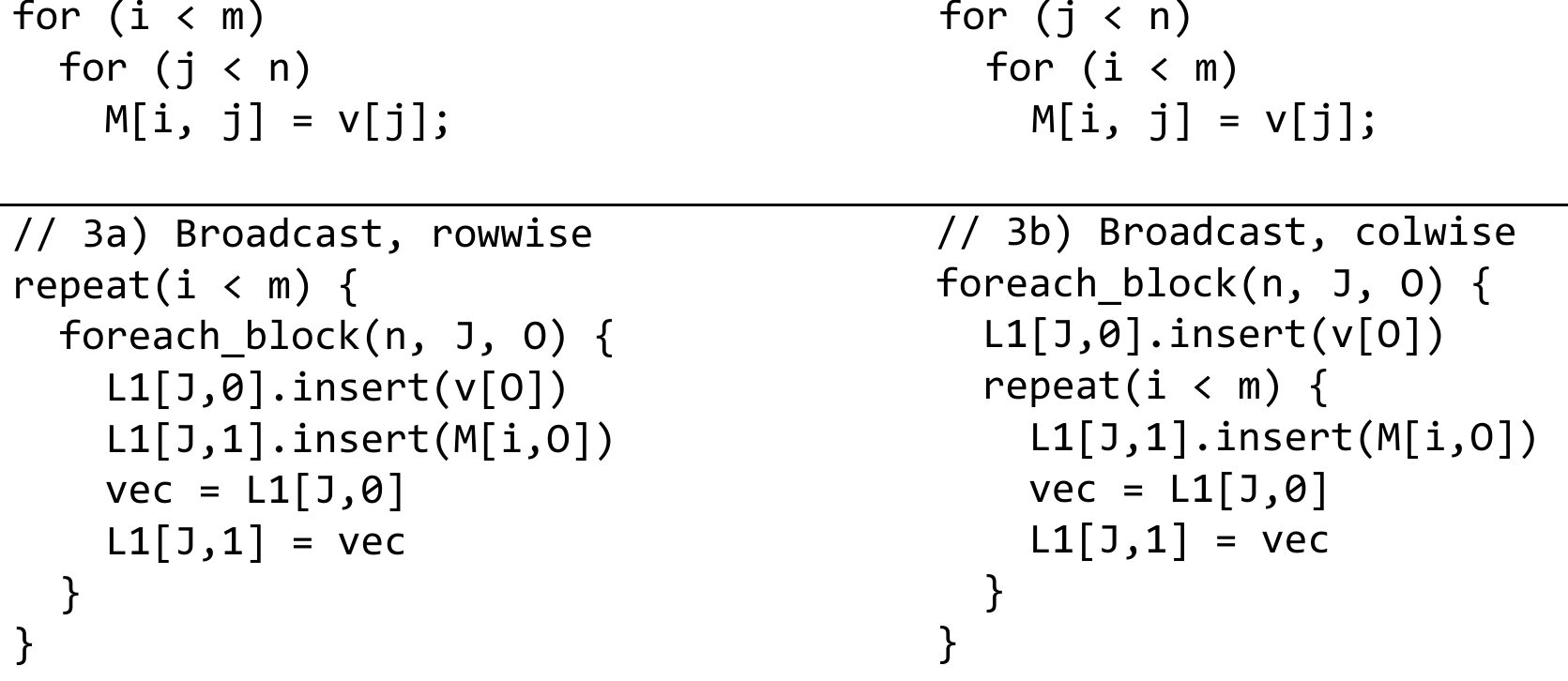}
  \caption{\label{fig:broadcast-reuse-src} Two broadcast kernels and their
    implementation in a hypothetical transparent programming model.}
\end{figure}

The next example is a broadcast kernel, two versions of which are listed in the
top of \Cref{fig:broadcast-reuse-src}. Both kernels broadcast a vector
of data into the rows of a matrix, but their loop order is reversed. We will
assume that the matrix is stored with a row-major layout.

\subsubsection{Data Reuse}

We now consider a programming model which exposes how L1 cache data is reused.
Again, we let \texttt{L} denote the number of elements in a cache line. We
extend the programming model from the previous example with two constructs. The
first is \texttt{foreach\_block(n, index, offset) body}, which is used to loop
over cache block indices in a vector and the corresponding cache set indices.
Specifically, the new loop construct executes \texttt{body}
$\frac{\text{n}}{\text{L}}$ times, with \texttt{offset} ranging from 0 to
\texttt{n - L} in steps of \texttt{L} and \texttt{index} bound to \texttt{offset
  \% L}. The second construct is \texttt{L1[set,way]}, which refers to a logical
cache line by indexing the \texttt{L1} object with logical set and way
indices\footnote{For direct-mapped or fully-associative caches, one index would
  suffice.}. These index expressions can be used as CPU instruction operands to
load and store a vector of data.

We also change how cache blocks are inserted into L1 from L2 compared to the
example in \Cref{sec:aos-vs.-soa}. Previously, this operation was performed with
the construct \texttt{L1.insert()}. In this example, we will instead use
\texttt{L1[set,way].insert()}, logically inserting the block at the cache line
referred to by \texttt{L1[set,way]}. Arbitrary variables can no longer hold the
result of L1 insert operations. The intention is to make the programmer relate
L1 use to its size, thereby increasing awareness of capacity misses. Data reuse
is context-dependent, and an implementation of such a model would also have to
rely on semantic analyses to assert that resident cache blocks cannot be
inserted. Note that the analyses need not be precise, as imperfections only
degrade performance transparency and not correctness.

\Cref{fig:broadcast-reuse-src} demonstrates how this programming model can be
used to describe the two example kernels. For simplicity, we assume that the
size of the source vector is a multiple of the vector length and greater than
half the L1 cache capacity. If so, the row-wise traversal can be described as in
code section 3a). At some point during the inner loop, source and vector data
which is loaded will begin evicting previously loaded source vector blocks. The
effect is that each block must be re-inserted into L1 when new rows are
traversed. In contrast, the inner loop of code section 3b) uses a single block
from the source vector, which due to temporal locality will not be evicted by
the destination blocks. Therefore, the vector source blocks are only inserted
once per outer-loop iteration.

\section{Performance Models}
\label{sec:performance-models}

We now consider how we can model the performance from these programs using
attribute grammars. Our objective is to define an attribute $c$ on nodes in the
grammar, representing the execution time in cycles as a symbolic formula. For
non-terminals which represent variables in the grammar, we use the attribute $n$
to denote the symbolic variable names. We use the symbolic constants
$T_{insert}$, $T_{load}$, $T_{store}$, $T_{rep}$, and $T_{gp}$ to denote the
execution time of inserting data from L2 cache into the L1 cache, read and write
access to data in the L1 cache, loop repetition overhead, and
\texttt{gather\_partial()}. The examples are kept simple by ignoring the cost of
address generation, which is valid if addresses can be formed directly from the
loop counter: if true, the address generation cost will be included in
$T_{rep}$.

The grammars are designed to be unambiguous for the examples in question, but
for simplicity we have not designed them with the objective of being efficiently
parsable. Note also that the grammars only demonstrate that the programming
model constructs are amenable to transparent performance predictions when used
in a context similar to our examples. If the constructs were to be included in a
more complex programming model, they might need to be adapted to avoid
ambiguity.

We use the following notational convention: capitalized words in italics are
non-terminals, lower-case words in bold face are terminal symbols, and
characters not in italics are terminals. Subscripts are used to separate
different occurrences of a non-terminal in a production. The terminal symbols
represent a string value with a certain meaning; this string value is assumed to
be available as the attribute \textit{string}.

\subsection{AoS vs. SoA}
\label{sec:aos-soa-perfmod}

\begin{table}[t]
  \centering
  \caption{Performance Prediction as an Attribute Grammar for Examples in
    \Cref{sec:aos-vs.-soa}.}
  \begin{tabular}{LR}
    
    \toprule
  \text{Productions} & \text{Semantic Rules} \\
    \midrule
  Prog \rightarrow Stmts & c(Prog) = c(Stmts) \\
  Stmts_1 \rightarrow Stmt \  Stmts_2 & c(Stmts_1) = c(Stmt) + c(Stmts_2)\\
  Stmts \rightarrow Stmt & c(Stmts) = c(Stmt) \\
  Stmt \rightarrow R & c(Stmt) = c(R) \\
  Stmt \rightarrow Insert & c(Stmt) = c(Insert)\\
  Stmt \rightarrow LoadLine & c(Stmt) = c(LoadLine) \\
  Stmt \rightarrow LoadElem & c(Stmt) = c(LoadElem) \\
  Stmt \rightarrow Store & c(Stmt) = c(Store) \\
  Stmt \rightarrow G & c(Stmt) = c(G) \\
  R \rightarrow \text{repeat(i < } V_1 \text{, step =} V_2\text{)} \  B &  c(R) =  \frac{n\left(V_1\right)}{n\left(V_2\right)}\left(c\left( B \right) + T_{rep} \right) \\
  R \rightarrow \text{repeat(i < } V \text{)} \  B &  c(R) =  n\left(V\right)\left(c\left( B \right) + T_{rep} \right) \\
  B \rightarrow \text{\{}\, Stmts \, \text{\}} & c(B) = c(Stmts)\\
  LoadLine \rightarrow V \text{ = } Deref & c(LoadLine) = T_{load} \\
  LoadElem \rightarrow V \text{ = } Index & c(LoadLine) = T_{load} \\
  LoadElem \rightarrow V_1 \text{ = } V_2 \text{.at(} E \text{)} & c(LoadElem) = T_{load} \\
  G \rightarrow V_1 \text{ = gather\_partial(}V_2 \text{, } V_3 \text{)} & c(G) = T_{gp} \\
  Insert \rightarrow V \text{ = L1.insert(} Index \text{)} & c(Insert) = T_{insert}\\
  Insert \rightarrow Index_1 \text{ = L1.insert(} Index_2 \text{)} & c(Insert) = T_{insert}\\
  StoreElem \rightarrow Index \text{ = } V & c(Store) = T_{store} \\
  Store \rightarrow Deref \text{ = } V & c(Store) = T_{store} \\
  V \rightarrow \text{\bf var} & n(V) = string(\text{\bf var}) \\
  Index \rightarrow V_1 \text{[} E \text{]} & \\
  Deref \rightarrow \text{*}V & \\
    E \rightarrow V & \\
    E \rightarrow E + E & \\
    E \rightarrow E * E & \\
    \bottomrule
  \end{tabular}
  \label{tab:aos-soa-v1-grammar}
\end{table}

\Cref{tab:aos-soa-v1-grammar} presents an attribute grammar which produces
performance prediction formulae for the examples in \Cref{sec:aos-vs.-soa}. For
the scalar implementations in \Cref{fig:aos-vs-soa-scalar-src}, the grammar
gives the following performance models:
 
\begin{empheq}[left=\Rightarrow]{align*}
  T_{1a} =&\frac{\text{n}}{\text{L}}\left( T_i + \text{S}\left( T_i + T_{rep} \right) + \text{L} \left( T_{load} + T_{store} + T_{rep} \right) \right) \\
  T_{1b} =&\frac{\text{n}}{\text{L}}\left( T_i + T_i + \text{L}\left( T_{load} + T_{store} + T_{rep} \right) \right) \\
  T_{1a}=&\frac{\text{n}}{\text{L}}\left( \left( 1 + \text{S} \right) T_i + ST_{rep}  \right) + \text{n} \left( T_{load} + T_{store} + T_{rep} \right) \\
  T_{1b}=&\frac{\text{n}}{\text{L}}\left( 2T_i \right) + \text{n}\left( T_{load} + T_{store} + T_{rep} \right)
\end{empheq}

The performance models reflect the source code: in 1a) the \texttt{L1.insert}
call is repeated \texttt{S} times for the source vector, whereas there is only
one such call in 1b).\footnote{The extra loop overhead in model $T_{1a}$ is an
  imperfection.} In other words, it is visible to the programmer that the AoS
version has a larger memory footprint than the SoA version, leading to a
run-time increase proportional to the size of the struct.

Applied to the vectorized implementations in \Cref{fig:aos-vs-soa-vec-src}, the
grammar in \Cref{tab:aos-soa-v1-grammar} produces the following prediction
formulae:
\begin{empheq}[left=\Rightarrow]{align*}
   T_{2a} =& \frac{\text{n}}{\text{L}}\left( T_i + \text{S}  \left( T_i + T_{gp}
     + T_{rep} \right) + T_{store} + T_{rep} \right) \\
   T_{2b} =& \frac{\text{n}}{\text{L}}\left( T_i + T_i + T_{load} + T_{store} +
     T_{rep} \right) \\
   T_{2a} =& \frac{\text{n}}{\text{L}}\left( \left( 1 + \text{S} \right)T_i + S\left( T_{gp} + T_{rep} \right) + T_{store} + T_{rep} \right) \\
   T_{2b} =& \frac{\text{n}}{\text{L}}\left( 2T_i + T_{load} + T_{store} + T_{rep} \right)
 \end{empheq}

    


 These performance models reveal two important considerations. First, the use of
 vector gather instructions does not alleviate the larger memory footprint of
 the AoS layout: it is still necessary to insert \texttt{S} lines of source data
 per line of destination data. Thus, we expect the SoA version to be faster even
 if the AoS code is vectorized. Second, on the Xeon Phi coprocessor the
 instruction count reduction from \texttt{gather()}-based vectorization is
 inversely proportional to the stride since the \texttt{gather\_partial()} call
 is also repeated \texttt{S} times. This means that we expect the benefit of
 vectorization using \texttt{gather()} to decline with increasing strides.

\subsection{Broadcast}
\label{sec:broadcast-perfmod}

\Cref{tab:broadcast-grammar} presents a grammar which models the performance of
the examples in \Cref{sec:broadcast}. The grammar assumes that the symbol
\texttt{L} can be used to refer to the number of elements in a cache line in the
production of \textit{F}: since this is type-specific, the grammar will only
work if all types have the same size which lets \texttt{L} be a constant. A more
advanced programming model would have to use separate \texttt{foreach\_block()}
versions for each type-size to instead select an appropriate value in the
performance model.

 \begin{table}[t]
  \centering
  \caption{Performance Prediction Grammar for Examples in
    \Cref{sec:broadcast}.}
  \begin{tabular}{LR}
    
    \toprule
    \text{Productions} & \text{Semantic Rules} \\
    \midrule
    Prog \rightarrow Stmts & c(Prog) = c(Stmts) \\
    Stmts_1 \rightarrow Stmt \  Stmts_2 & c(Stmts_1) = c(Stmt) + c(Stmts_2)\\
    Stmts \rightarrow Stmt & c(Stmts) = c(Stmt) \\
    Stmt \rightarrow R & c(Stmt) = c(R) \\
    Stmt \rightarrow Insert & c(Stmt) = c(Insert)\\
    Stmt \rightarrow LoadLine & c(Stmt) = c(LoadLine) \\
    Stmt \rightarrow Store & c(Stmt) = c(Store) \\
    Stmt \rightarrow F & c(Stmt) = c(F) \\
    R \rightarrow \text{repeat(i < } V \text{)} \  B &  c(R) =  n\left(V\right)\left(c\left( B \right) + T_{rep} \right) \\
    F \rightarrow \text{foreach\_block(} V_1 \text{, } V_2 \text{, } V_3\text{)} \  B &  c(F) =  \frac{n\left(V_1\right)}{\text{L}}\left(c\left( B \right) + T_{rep} \right) \\
  B \rightarrow \text{\{}\, Stmts \, \text{\}} & c(B) = c(Stmts)\\
  LoadLine \rightarrow V \text{ = } LineRef & c(LoadLine) = T_{load} \\
  Store \rightarrow LineRef \text{ = } V & c(Store) = T_{store} \\
  Insert \rightarrow LineRef \text{.insert(} Index \text{)} & c(Insert) = T_{insert}\\
    LineRef \rightarrow \text{L1[} V \text{, } \textbf{int} \text{]} & \\
    \bottomrule
  \end{tabular}
  \label{tab:broadcast-grammar}
\end{table}

The code leads to the following performance predictions:

\begin{empheq}[left=\Rightarrow]{align*}
  T_{3a} =&&m\left(\frac{n}{L} \left(  T_i + T_i + T_l + T_s + T_{rep} \right) +
  T_{rep} \right) & \\
  T_{3b} =&&\frac{n}{L}\left(T_i + m\left(T_i + T_l + T_s + T_{rep}\right) +
    T_{rep} \right) & \\
  T_{3a} =&&\frac{nm}{L}\left(2T_i + T_l + T_s + \left( 1 + \frac{L}{n} \right)T_{rep}\right) & \\
  T_{3b} =&&\frac{nm}{L}\left(\left(1 + \frac{1}{m})\right)T_i + T_l + T_s +
    \left( 1 + \frac{1}{m} \right) T_{rep} \right) &
\end{empheq}
%

The performance models reflect the increased data reuse in code section 3b),
which leads to fewer cache line inserts compared to 3a) depending on the number
of rows in the array. There is also a difference in loop overhead: the inner
loop overhead will occur the same number of times, but the outer loop overhead
depends on which loop is outermost.

\section{Evaluation Method}
\label{sec:evaluation-method}

To evaluate the accuracy of the performance models, we use the execution times
listed in \Cref{tab:runtime-cost} to predict program performance. Actual
performance is measured from running C++ implementations presented in
Listing~\ref{lst:aos-soa-impl-v1}, which are designed to correspond to the
implementation alternatives of
\Cref{fig:aos-vs-soa-scalar-src,fig:aos-vs-soa-vec-src,fig:broadcast-reuse-src}.
The code is compiled with Intel's \texttt{icpc} compiler, version 17. We use the
data type \texttt{float}, which is four bytes on Intel Xeon Phi. As the platform
uses 512-bit vectors, the number of elements in a vector is \texttt{16}. These
C++ implementations could be translated in many different ways. To ensure that
the translations match the implementation alternatives specified in
\Cref{fig:aos-vs-soa-scalar-src}, we use non-const pointer parameters to prevent
vectorization in examples 1a) and 1b) due to potential aliasing. We also use the
flag \texttt{-fno-builtin} to prevent the SoA implementations, 1b) and 2b), from
being implemented as a call to \texttt{memcpy}. For implementations 2a) to 3b),
we use manual vectorization.

\begin{table}[t]
  \centering
  \caption{Execution Time of Operations}
  \begin{tabular}{llr}
    \toprule
    Symbol & Operation & Cycles \\
    \midrule
    \textit{T\textsubscript{insert}} & Insert data from the L2 cache into the L1 cache & 20 \\
    \textit{T\textsubscript{load}} & Load data from L1 cache into a register & 2 \\
    \textit{T\textsubscript{store}} & Store data from a register into L1 cache & 2 \\
    \textit{T\textsubscript{rep}} & Loop repetition overhead & 2 \\
    \textit{T\textsubscript{gp}} & Gather data from one L1 cache line & 4 \\
    \bottomrule
  \end{tabular}
  \label{tab:runtime-cost}
\end{table}

\begin{listing}[p]
\begin{subfigure}[b]{.5\linewidth}
\begin{minted}[fontsize=\footnotesize]{c++}
struct S_aos {
  float x;
  char pad[(STRUCT_SIZE - 1)*
           sizeof(float)];
};
// 1a)
void AoS(float*& d,
         S_aos*& s,
         std::size_t n) {
  __assume_aligned(d, 64);
  __assume_aligned(s, 64);
  __assume(n % 16 == 0);
  std::size_t i;
  for (i = 0; i < n; ++i) {
    d[i] = s[i].x;
  }
}
\end{minted}
\end{subfigure}
\hspace{0.5em}
\begin{subfigure}[b]{.5\linewidth}
\begin{minted}[fontsize=\footnotesize]{c++}
struct S_soa {
  float* x;
};
// 1b)
void SoA(float*& d,
         S_soa& s,
         std::size_t n) {
  __assume_aligned(d, 64);
  __assume_aligned(s.x, 64);
  __assume(n % 16 == 0);
  std::size_t i;
  for (i = 0; i < n; ++i) {
    d[i] = s.x[i];
  }
}
\end{minted}
\end{subfigure}

\begin{subfigure}[b]{.5\linewidth}
\begin{minted}[fontsize=\footnotesize]{c++}
// 2a)
void AoS_gather(float*& d,
                S_aos*& s,
                std::size_t n) {
  __assume_aligned(d, 64);
  __assume_aligned(s, 64);
  __assume(n % 16 == 0);
  __declspec(aligned(64))
      int indices [] = {
    0,1*STRUCT_SIZE,...,15*STRUCT_SIZE
  };
  __m512i idxvec =
    _mm512_load_epi32(indices);
  auto scale = 1; 
  std::size_t i;
  for (i = 0; i < n; i += VL) {
    __m512 vec = _mm512_i32gather_ps(
         idxvec, &s[i], scale);
    _mm512_store_ps(&d[i], vec);
  }
}
\end{minted}
\end{subfigure}
\hspace{0.5em}
\begin{subfigure}[b]{.5\linewidth}
\begin{minted}[fontsize=\footnotesize]{c++}
// 2b)
void SoA_vec(float*& d,
             S_soa& s,
             std::size_t n) {
  __assume_aligned(d, 64);
  __assume_aligned(s.x, 64);
  __assume(n % 16 == 0);
  std::size_t i;
  for (i = 0; i < n; ++i) {
    __m512i vec =
       _mm512_load_epi64(&s.x[i]);
    _mm512_store_epi64(&d[i], vec);
  }
}
\end{minted}
\end{subfigure}

\begin{subfigure}[b]{.5\linewidth}
\begin{minted}[fontsize=\footnotesize]{c++}
// 3a)
void rowwise(const float* v,
             float* M,
             std::size_t n,
             std::size_t m) {
  for (auto i = 0u; i < m; ++i) {
    for (auto j = 0u; j < n;
         j += 16, M += 16) {
      __m512 vec =
          _mm512_load_ps(&v[j]);
      _mm512_store_ps(M, vec);
    }
  }
}
\end{minted}
\end{subfigure}
\hspace{0.5em}
\begin{subfigure}[b]{.5\linewidth}
\begin{minted}[fontsize=\footnotesize]{c++}
// 3b)
void colwise(const float* v,
             float* M,
             std::size_t n,
             std::size_t m)
  float* M_orig = M;
  for (auto j = 0u; j < n; j += 16) {
    M = &M_orig[j];
    for (auto i = 0u; i < m; i += n) {
      __m512 vec = _mm512_load_ps(&v[j]);
      _mm512_store_ps(&M[i], vec);
    }
  }
}
\end{minted}
\end{subfigure}

\caption{\label{lst:aos-soa-impl-v1} C++ implementations of the examples in
  \Cref{fig:aos-vs-soa-scalar-src,fig:aos-vs-soa-vec-src,fig:broadcast-reuse-src}.}
\end{listing}

There are two parameters in the Aos vs. SoA example: the size of the struct, and
the number of elements in the arrays. The broadcast example also has two
parameters: the number of rows, and the source vector size. We measure and
predict performance for the set of parameter values listed in
\Cref{tab:param-space-ex1}.

\begin{table}[t]
  \centering
  \caption{Parameter Space Used to Test Listing~\ref{lst:aos-soa-impl-v1}.}
  \begin{tabular}{lr}
    \toprule
    Parameter & Values \\
    \midrule
    Struct Size & 1, 2, ..., 16  \\
    Num. Elements & $2^6$, $2^7$, ..., $2^{10}$ \\
    Rows & 1, 2, ..., 10 \\
    Source Vector Size [KiB] & 20, 24, ..., 48 \\
    \bottomrule
  \end{tabular}
  \label{tab:param-space-ex1}
\end{table}

Accuracy metrics are in general obtained by comparing predictions $\hat{y}$ with
the true performance distribution $y_i$. We measure accuracy as median absolute
percentage error (MdAPE) \cite{Hyndman2006}:

\[
    \text{MdAPE} = median \left(  \frac{\lvert y_i - \hat{y} \rvert}{y_i} \right)
\]

This accuracy measure may be interpreted as follows: for each performance
measurement the prediction is $X \%$ wrong, and the MdAPE is the median of these
percentage-errors. This is independent of the scale of the true performance
values, and may therefore be compared across loads. The median is less sensitive
to outliers than the mean, which makes it more lenient towards
rare-but-detrimental performance events such as page faults. Absolute value
ensures that symmetric variability in the true performance is not ignored, and
is less sensitive to outliers than alternatives such as squared error. The MdAPE
is only appropriate if the observed values are not close to zero
\cite{Hyndman2006}; we therefore measure execution time as clock cycles.

\section{Results}
\label{sec:results}

\subsection{AoS vs. SoA}
\label{sec:aos-soa-results}

We first report on the ability to use the predictions to assess which
implementation will perform better. The best-performing method is determined by
comparing the median execution times.

Comparing 1a) with 1b), the performance model always predicts that the SoA
implementation is faster. This is consistent with the measurements in all cases
except when the struct size is 1. In this case, the two implementations ought to
be equal. We even observe that the median execution time of the AoS
implementation is 5 cycles lower, which we attribute to a difference in code
generation. Comparing 2a) with 2b), the performance model again predicts that
the SoA implementation will always be faster. This is consistent with all
measurements: the overhead of the gather instruction makes it perform worse than
the SoA implementation even when the struct size is 1.

\begin{figure}[t]

\begin{subfigure}[t]{.5\linewidth}
  \centering
  \includegraphics[scale=0.95]{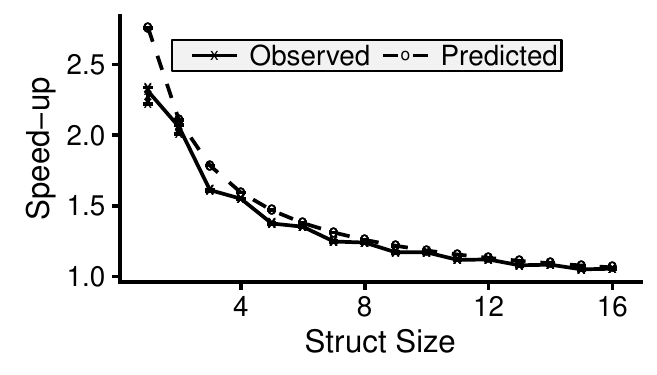}
  \caption{\label{fig:relperf-aos-vec-value} Speed-up of 2a) over 1a).}
\end{subfigure}
\begin{subfigure}[t]{.5\linewidth}
  \centering
  \includegraphics[scale=0.95]{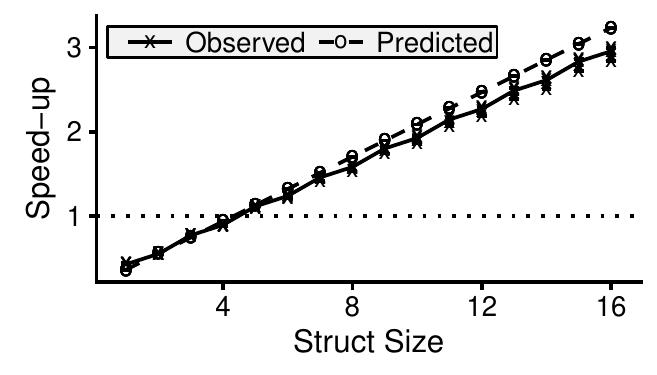}
  \caption{\label{fig:relperf-soa-value} Speed-up of 1b) over 2a).}
\end{subfigure}
  \caption{\label{fig:relperfs} Observed and predicted speed-ups
    across implementations.}
\end{figure}

Comparing 1a) to 2a) to
determine the usefulness of vectorization in the AoS case, we correctly predict
that it is always beneficial to use vectorization.
\Cref{fig:relperf-aos-vec-value} plots the predicted and observed usefulness of
vectorization, measured as the relative runtime of the non-vectorized
implementation to the vectorized implementation. The predicted trend is
followed, but every second data point has larger inaccuracies. We attribute this
to the code generated for the \texttt{gather()} intrinsic: the loop is unrolled
twice, leading to more complex branching behaviour when the number of iterations
required is not a multiple of two.

Finally, comparing 1b) to 2a) to determine the usefulness of gather-based
vectorization versus the usefulness of the SoA layout, we correctly determine
that SoA is more impactful once the struct size exceeds 4.
\Cref{fig:relperf-soa-value} plots the predicted and observed relative
performance of implementation 1b) over implementation 2a), which are largely similar.

\begin{figure}[t]
  \centering
  \includegraphics[scale=0.95]{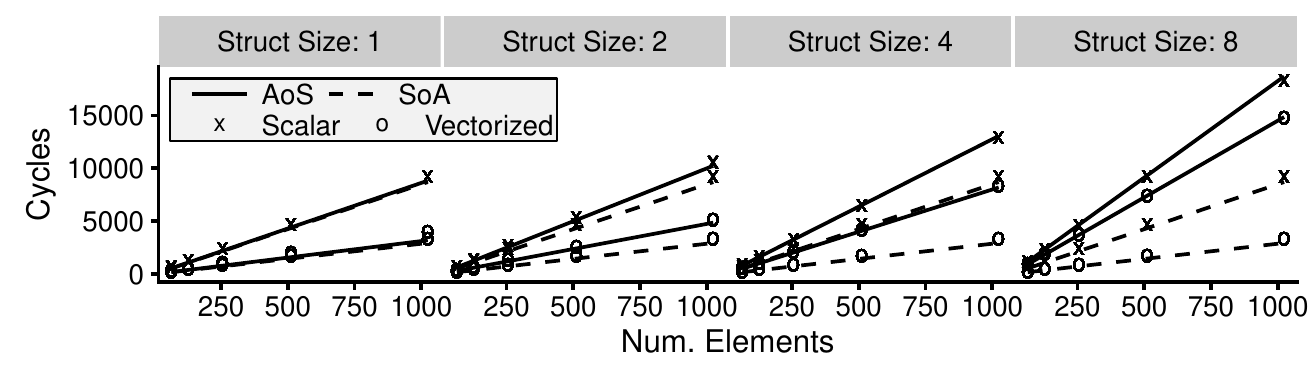}
  \caption{\label{fig:aos-soa-all} Performance of the code in
    \Cref{fig:aos-vs-soa-scalar-src,fig:aos-vs-soa-vec-src} for a subset of the
    parameters in \Cref{tab:param-space-ex1}. Points are observations, and lines
    are predictions.}
\end{figure}

\Cref{fig:aos-soa-all} plots predictions along with observations for a subset of
the parameters tested, illustrating that the programming models in
\Cref{fig:aos-vs-soa-scalar-src} give an accurate representation of processor
behaviour.

\begin{figure}[t]
  \centering
  \includegraphics[scale=0.95]{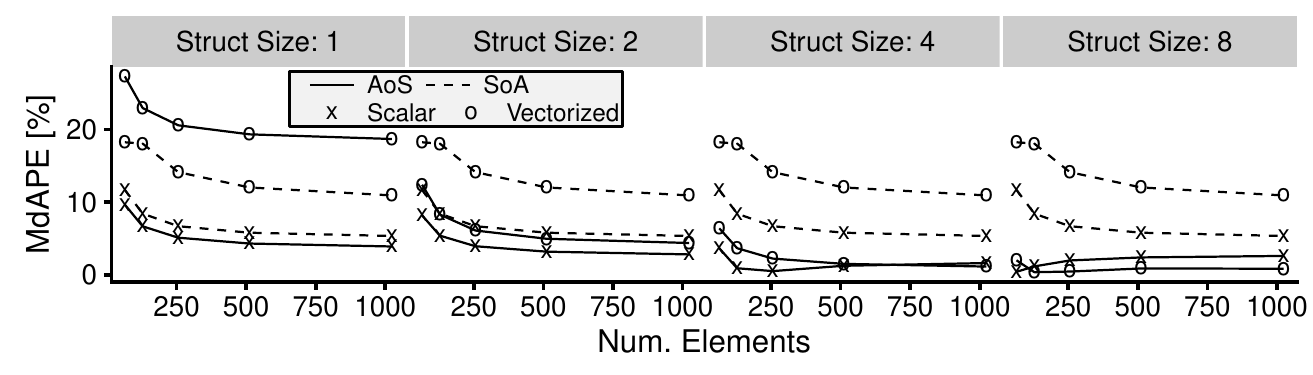}
  \caption{\label{fig:aos-soa-all-acc} The accuracy of predictions
    in \Cref{fig:aos-soa-all}.}
\end{figure}

\begin{figure}[t]
  \centering
  \includegraphics[scale=0.95]{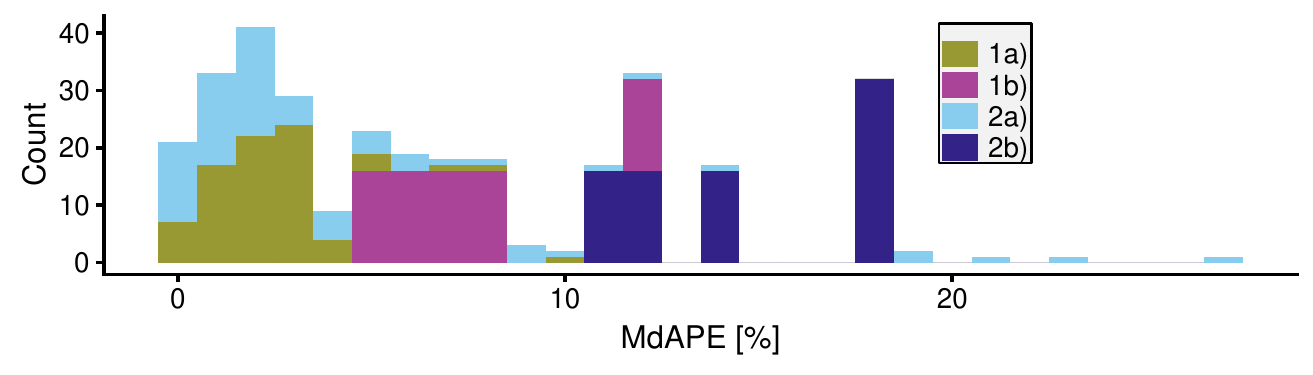}
  \caption{\label{fig:aos-soa-scalar-acc} The distribution of accuracy
    measurements for the example in \Cref{sec:aos-vs.-soa} using all the
    parameter values in \Cref{tab:param-space-ex1}.}
\end{figure}

\Cref{fig:aos-soa-all-acc} plots the prediction accuracy for the results in
\Cref{fig:aos-soa-all}. Accuracy is generally poorer for shorter vectors in the
SoA case, and for smaller strides in the AoS case. The complete distribution of
accuracy measurements using the parameter set in \Cref{tab:param-space-ex1} for
all implementations is depicted in \Cref{fig:aos-soa-scalar-acc}. The predictions for
the AoS implementations are in general better than the predictions for the SoA
implementation.


\subsection{Broadcast}
\label{sec:broadcast-results}

\begin{figure}[t]

  \centering
  \includegraphics[scale=0.95]{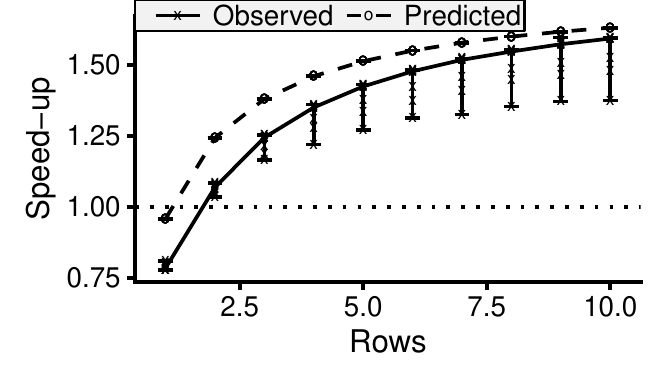}
  \caption{\label{fig:relperf-broadcast} Observed and predicted speed-ups
    of 3b) over 3a).}
\end{figure}

For the example in \Cref{sec:broadcast}, the performance models predict that the
column-wise implementation will perform better except when the number of rows
equals 1: for this parameter setting, the increased loop overhead will tip the
scales in favour of the row-wise implementation. This prediction is consistent
with the observed behaviour. However, the magnitude of the predicted speed-up is
not equal to the observed speed-up when the number of rows is low, as figure
\Cref{fig:relperf-broadcast} illustrates. This is caused by two effects. First,
our model for the \texttt{repeat} abstraction does not include overhead for
initialization of loop counters. Our model therefore underestimates the
execution time of implementation 3b), which has a higher outer-loop than
inner-loop count. Second, different code is generated for the two
implementations in \Cref{lst:aos-soa-impl-v1}. In particular, there is a RAW
hazard between the load and store instructions which is not hidden in
implementation 3b) but which is hidden by loop overhead instructions in the code
generated for 3a). These effects are diminished as the number of rows increase,
where the data reuse effect increasingly dominate the performance.

\begin{figure}[t]
  \centering
  \includegraphics[scale=0.95]{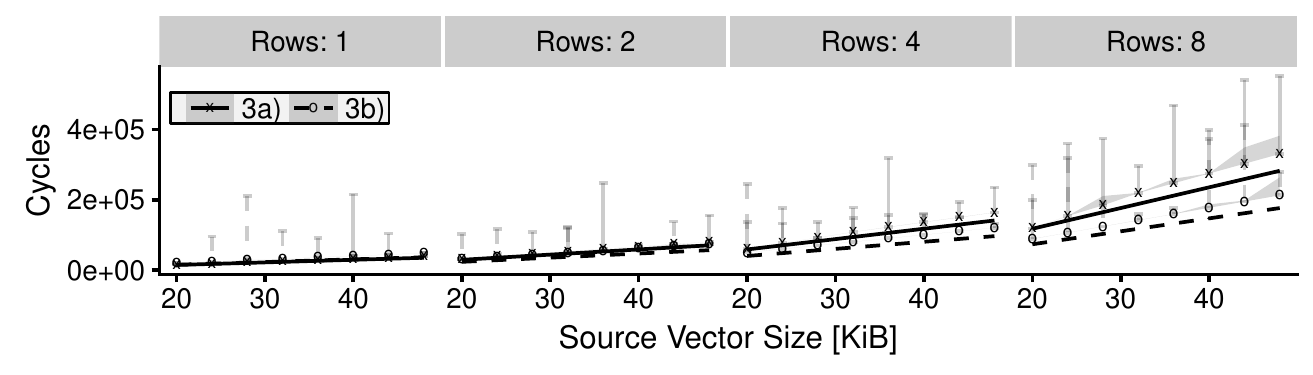}
  \caption{\label{fig:rowwise-vs-colwise} Performance of the code in
    \Cref{fig:broadcast-reuse-src}. Points are the median of observed
    performance, the light-gray ribbon marks the 97-percent confidence interval,
    and error bars denote the span of observations. The lines are predictions.}
\end{figure}

\begin{figure}[t]
  \centering
  \includegraphics[scale=0.95]{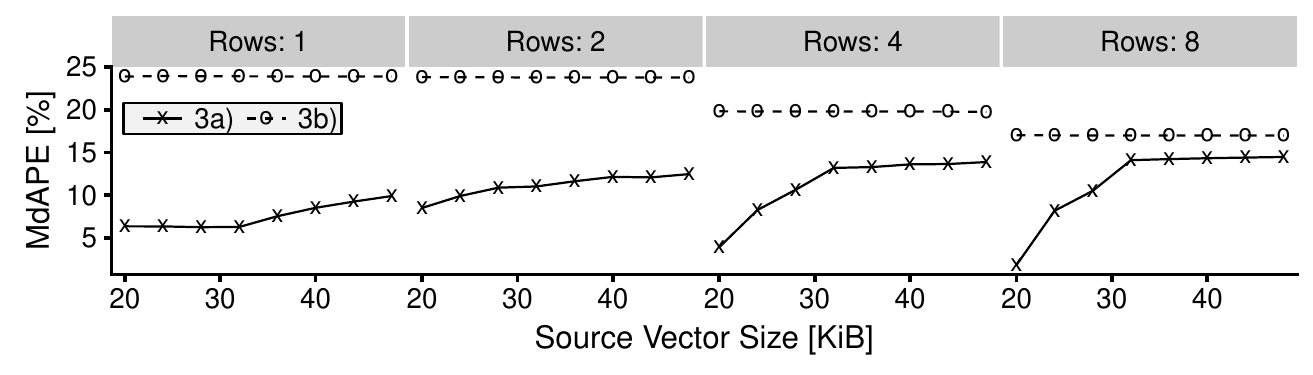}
  \caption{\label{fig:rowwise-vs-colwise-acc} Prediction accuracy for the test
    cases illustrated in \Cref{fig:rowwise-vs-colwise}.}
\end{figure}

\begin{figure}[t]
  \centering
  \includegraphics[scale=0.95]{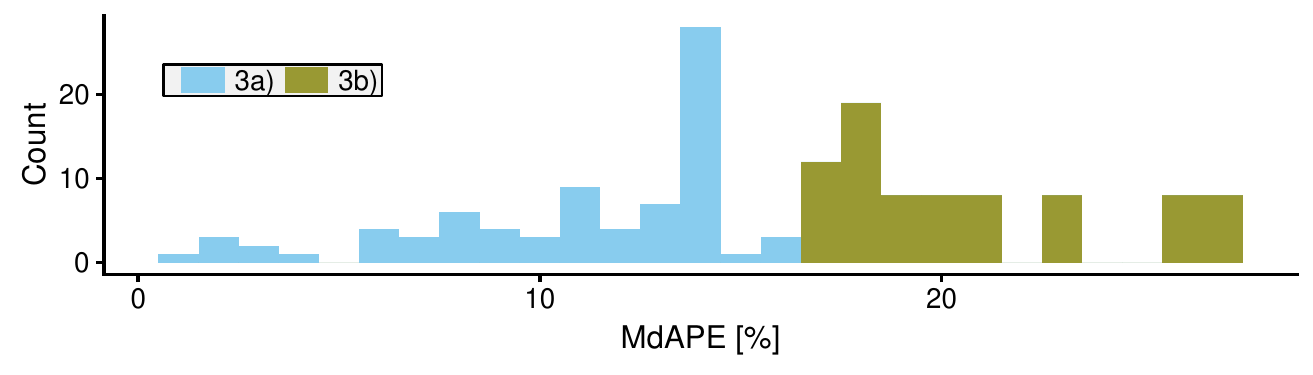}
  \caption{\label{fig:broadcast-acc} The distribution of accuracy measurements
    for the example in \Cref{sec:broadcast} using all the parameter values in
    \Cref{tab:param-space-ex1}.}
\end{figure}

\Cref{fig:rowwise-vs-colwise} illustrates how the code in
\Cref{fig:broadcast-reuse-src} and its performance model reflect observed
behaviour. \Cref{fig:rowwise-vs-colwise-acc} plots the corresponding prediction
accuracy measurements, and \Cref{fig:broadcast-acc} plots the accuracy for all
test cases listed in \Cref{tab:param-space-ex1}. The accuracy is in general
lower than what we observe for the examples in \Cref{sec:aos-soa-results}.

\section{Discussion}
\label{sec:discussion}

\subsection{Implementation}
\label{sec:implementation}

A natural question is how the hypothetical programming models may be
implemented. For instance, the semantics of the \texttt{L1.insert()} operation
is not commonly accessible through a specific processor instruction. Such an
instruction is unnecessary, however: the only purpose of the
\texttt{L1.insert()} operation is to document the fact that a load from L2 cache
to L1 cache will occur. All \texttt{L1.insert()} occurrences can therefore be
removed during compilation, as load instructions must also be generated for the
corresponding use of the cache line variable. This will of course require that
the compiler keeps track of which \texttt{L1.insert()} argument corresponds to a
given cache line variable, which further implies that the cache line variables
must be immutable. Another issue is how we may assure the correctness of the
\texttt{L1.insert()} operation, which depends both on the dynamic application
behaviour and on the replacement policy of the target machine. The compiler
would have to rely on semantic analyses such as \cite{Bao2018, Theiling2000,
  Guan2014} to deduce whether the cache use as expressed will be representative
of actual behaviour. We are looking to address these questions of implementation
in future work.

\subsection{Limited accuracy}
\label{sec:unimpr-accur}

Although the performance models in most cases are able to predict the
best-performing method, the accuracy measurements are at times low. Some of this
may be attributed to a methodical shortcoming: rather than using C++
implementations, it would have been better to use hand-written assembly programs
to get complete control of the loads. Even so, the fact that the generated code
has performance up to 25 \% away from our expectations reveals that performance
may be significantly affected by implementation details beneath the level
exposed by our example programming models. This means that we would need to
expose more details to increase performance transparency. In particular:

\begin{itemize}

\item The processor front-end may matter. For instance, the Xeon Phi is a
  super-scalar architecture with two pipelines \cite{Phi2012}, but the ability
  to issue two instructions depends on instruction pair compatibility.

\item RAW hazards may affect the performance of code, and without exposing
  hazards one cannot deduce the performance of code even if the instruction
  schedule is known.

\item The higher level abstractions such as loops, address generation, and
  \texttt{gather\_partial()} hide implementation details which affect program
  execution time.

\end{itemize}

Including these effects would complicate the programming model, and the effects
can be derived from the compiled binary code. An alternative solution is to let
the programmer use the less transparent programming model, and have a
performance modelling tool processes the compiled binary and produce both a
performance estimate and a corresponding implementation in a more transparent
programming model. Such alternative uses of the performance-transparent
programming models can be used to balance the trade-off between conciseness and
performance transparency.





\section{Conclusion}
\label{sec:conclusion}

This report has shown some proof-of-concept examples of performance-transparent
programming models. We have demonstrated how programming model abstractions may
be used to reveal the memory footprint, vector unit utilization, and data reuse
of an application, and how performance models which adhere to our transparency
requirements can be created. The examples demonstrate that the approach suffices
to determine which among alternative implementations is preferable. The absolute
performance prediction accuracy varies between 0-25 \%, which indicates that
increased performance transparency may be possible by exposing even more details
of the processing model. In future work, we will seek to implement some of the
ideas presented in our example programming models.


\end{document}